\documentclass[aps,prd,twocolumn,showpacs,superscriptaddress,nofootinbib,preprintnumbers]{revtex4-1}

\usepackage{comment}
\usepackage{slashed}
\usepackage{latexsym}
\usepackage{graphics}
\usepackage{bm}
\usepackage{color}
\usepackage{amsmath}
\usepackage{amssymb}
\usepackage{mathrsfs}
\usepackage{graphicx}
\usepackage{slashed}
\usepackage{siunitx}
\usepackage{braket}
\usepackage{calligra}
\usepackage{caption}
\usepackage{subcaption}

\usepackage{caption}
\usepackage{subcaption}
\definecolor{linkcolor}{rgb}{0.7752941176470588, 0.22078431372549023, 0.2262745098039215}

\usepackage{hyperref}
\hypersetup{colorlinks=true,
linkcolor=linkcolor,
citecolor=linkcolor,
urlcolor=linkcolor,
linktocpage=true
}

\def\lsim{\mathrel{\rlap{\lower4pt\hbox{\hskip1pt$\sim$}}
    \raise1pt\hbox{$<$}}}                
\def\gsim{\mathrel{\rlap{\lower4pt\hbox{\hskip1pt$\sim$}}
    \raise1pt\hbox{$>$}}}                

\newcommand{\Ap}{A^\prime}

\newcommand{\be}{\begin{eqnarray}}
\newcommand{\ee}{\end{eqnarray}}

\newcommand{\benum}{\begin{enumerate}}
\newcommand{\eenum}{\end{enumerate}}
\newcommand{\bi}{\begin{itemize}}
\newcommand{\ei}{\end{itemize}}

\newcommand{\Eq}[1]{Eq.~(\ref{#1})}  
\newcommand{\brac}[2]{ \left( \frac{#1}{#2} \right) }

\definecolor{nicegreen}{rgb}{0.1,0.5,0.1}

\DeclareSIUnit\electronvolt{e\kern-.05em V}
\DeclareSIUnit\tonneyear{tonne-year}

\newcommand{\intd}[1]{\int \frac{d^4 #1}{(2\pi)^4}}
\makeatletter 
    
\renewcommand\onecolumngrid{
\do@columngrid{one}{\@ne}%
\def\set@footnotewidth{\onecolumngrid}
\def\footnoterule{\kern-6pt\hrule width 1.5in\kern6pt}%
}

\renewcommand\twocolumngrid{
        \def\footnoterule{
        \dimen@\skip\footins\divide\dimen@\thr@@
        \kern-\dimen@\hrule width.5in\kern\dimen@}
        \do@columngrid{mlt}{\tw@}
}%

\makeatother 

\begin{document}

\hfill{\small FERMILAB-PUB-22-670-T} 

\medskip
\title{Freezing In Vector Dark Matter Through Magnetic Dipole Interactions}

\author{Gordan Krnjaic} \email{krnjaicg@fnal.gov}
\affiliation{Theoretical Physics Department, Fermi National Accelerator Laboratory, Batavia, IL 60510}
\affiliation{Department of Astronomy and Astrophysics, University of Chicago, Chicago, IL 60637}
\affiliation{Kavli Institute for Cosmological Physics, University of Chicago, Chicago, IL 60637}

\author{Duncan Rocha} \email{drocha@uchicago.edu}
\affiliation{Department of Physics, University of Chicago, Chicago, IL 60637}

\author{Anastasia Sokolenko} \email{sokolenko@kicp.uchicago.edu}
\affiliation{Theoretical Physics Department, Fermi National Accelerator Laboratory, Batavia, IL 60510}
\affiliation{Kavli Institute for Cosmological Physics, University of Chicago, Chicago, IL 60637}

\date{\today}

\begin{abstract}
We study a simple model of vector dark matter that couples
to Standard Model particles via magnetic dipole interactions. In this scenario,  
the cosmological abundance arises through the freeze-in mechanism
and depends on the dipole coupling, the vector mass, and the reheat temperature.
To ensure cosmological metastability, the vector must be lighter
than the fermions to which it couples, but rare decays can still
produce observable 3$\gamma$ final states; two-body decays can also occur 
at one-loop with additional weak suppression, but are subdominant if the vector
couples mainly to light fermions. 
For sufficiently heavy vectors, induced kinetic mixing with the photon can also yield additional two body decays to lighter
fermions and predict indirect detection signals through
final state radiation. 
We explore the implications
of couplings to various flavors of visible particles and emphasize
leptophilic dipoles involving electrons, muons, and taus, which offer the most
promising indirect detection signatures through 3$\gamma$, $e^+ e^- \gamma$, and $\mu^+ \mu^- \gamma$ decay channels.
We also present constraints from current and past telescopes, and sensitivity projections for future missions including  e-ASTROGAM and AMEGO. 
\end{abstract}

\maketitle

\section{Introduction}
\label{sec:intro}

While the evidence for the existence dark matter (DM) is overwhelming,
its microscopic properties remain elusive (see \cite{Bertone:2016nfn} for a historical review). Since there are few clues about its non-gravitational
interactions, it is currently not known
how DM was produced in the early universe or when that 
production took place. Thus, there is great motivation
to identify and test all predictive mechanisms 
for this key epoch in the history of the universe. 

Cosmological ``freeze-in" is among the simplest and most predictive DM production mechanisms \cite{Dodelson:1993je,Hall:2009bx}. In this scenario, DM is initially
not present at reheating when the hot radiation bath is first 
established. Rather, 
its density builds up through ultra-feeble interactions with Standard Model (SM)
particles and production halts when this  process becomes Boltzmann suppressed. Since these reaction rates are sub-Hubble, the DM 
never equilibrates with the SM, so unlike freeze-out, there is no need to 
deplete the large thermal entropy  with additional DM annihilation when equilibrium is lost. Freeze-in production ends when the temperature of the universe cools below either the mass of either the DM or the SM species to which it couples, whichever is 
greater.

 It is well known that 
dark photons $\Ap$ can be produced via freeze-in through a kinetic mixing interaction with the SM photon \cite{Pospelov_2008}.  Since $\Ap$ are also unstable and decay through this same interaction,
only the $m_{\Ap} < 2 m_e$ mass range can be cosmologically
metastable to provide a DM candidate. In this range, the 
DM decays via $\Ap \to 3 \gamma$ reactions and predicts a late time
X-ray flux uniquely specified by the $\Ap$ mass, once the kinetic mixing parameter is fixed 
to obtain the observed DM abundance. However, this tight relationship
between abundance and flux has been used to sharply constrain this simple model with observations
of X-ray lines, extragalactic background light,
and direct detection via absorption \cite{Pospelov_2008,An:2014twa}. Collectively, these
probes have eliminated nearly all viable parameter space for vector DM produced through
freeze-in via kinetic mixing interactions. 

In this paper we generalize dark photon freeze-in to allow for the possibility that its main interaction with visible matter is 
a magnetic dipole coupling to charged fermions, instead of kinetic mixing with the photon. Since the dipole operator has mass dimension 5, 
the freeze-in abundance is sensitive to the reheat temperature. Thus, unlike kinetic mixing, 
there is a parametric separation between the production rate at early times and the decay
rate at late times; the former depends on the reheat temperature and the latter does not, so it is possible
to achieve the observed DM abundance with a much feebler coupling to SM particles and, thereby, open up viable 
parameter space for direct and indirect detection.

  \begin{figure}[t]
\center
\includegraphics[width=3.5in]{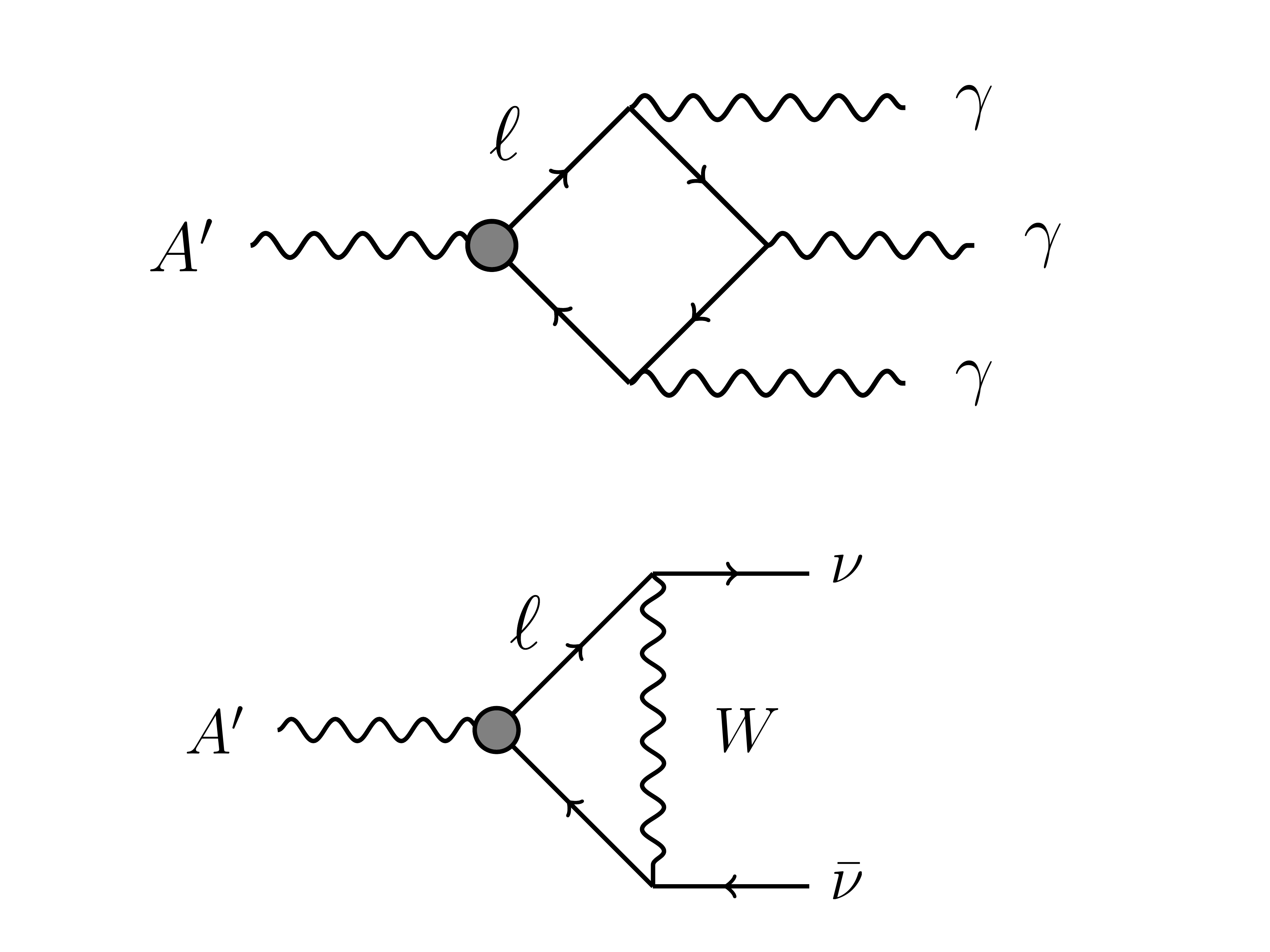}
\caption{Feynman diagrams representing the $\Ap \to 3\gamma$ (top) and
the $\Ap \to \bar \nu \nu$ decay (bottom) channels for $m_{\Ap} < 2m_\ell$. In both processes, the gray dot at the $\Ap$-$\ell$ vertex is the magnetic dipole interaction from \Eq{eq:lint} for charged leptons $\ell = e, \mu, \tau$. There are corresponding
diagrams involving quarks for which the electroweak loop yields decays to lighter
quark flavors instead of neutrinos. }
\label{fig:cartoon}
\end{figure}

This paper is organized as follows: in Sec. \ref{sec:theory} we describe the model, in Sec. \ref{sec:cosmo} we calculate the $\Ap$ abundance via freeze in, in Sec. \ref{sec:structure} we present structure formation limits, in Sec. \ref{sec:xray} we explore
the indirect detection constraints and future projections for this model,
and in Sec. \ref{sec:conclusion} we offer some concluding remarks.




\section{Theory Overview} \label{sec:theory}

\subsection{Model Description}
We extend the SM with a hidden $U(1)_H$ group with corresponding gauge boson $\Ap$, which doesn't couple to any SM particles through renormalizable operators. The  leading infrared interaction between $\Ap$ and SM particles is taken to be a magnetic dipole coupling
\be
\label{eq:lint}
{\cal L}_{\rm int} = \frac{d_f}{2}  F_{\mu\nu}^\prime \bar f \sigma^{\mu\nu} f ~,
 \ee
where $f$ is a charged SM fermion, $d_f$ is the corresponding magnetic dipole moment,
and $F^\prime_{\mu\nu}$ is the $\Ap$ field strength tensor.  
Such an interaction can arise if the $U(1)_H$ is unbroken at high energies and heavy
particles
charged appropriately under $U(1)_H$ and the SM are integrated out at low energies.\footnote{See Ref. \cite{Dobrescu_2005} 
for an explicit construction involving two-loop diagrams with virtual exchange of both $U(1)_H$ charged and SM charged
particles. In this example, it is important that the new states are not bifundamentals under the SM and the hidden
group so that kinetic mixing doesn't arise at lower loop order.} 
Since the magnetic dipole coupling is a dimension-5
operator, \Eq{eq:lint} is only 
valid at energy (or temperature) scales that satisfy $E \ll d_f^{-1}$.

If the $\Ap$ is initially massless, any potential
kinetic mixing between $U(1)_H$ and $U(1)_Y$ gauge bosons can be rotated away, so the operator in \Eq{eq:lint} can be the dominant interaction with SM particles \cite{Dobrescu_2005}. 
However, for $\Ap$ to be a viable dark matter candidate, it must acquire a
mass at some lower energy scale, at which point kinetic mixing of the form
$\frac{\epsilon}{2} F^{\mu\nu}F^\prime_{\mu\nu}$ 
 can arise from loops of SM particles through their dipole interactions, where
\be
\label{eq:kinmix}
\epsilon \sim \frac{e d_f m_f}{4 \pi^2} \approx  2 \times 10^{-15} \brac{m_f}{m_e}\brac{d_f \cdot \rm GeV}{10^{-8}}~,
\ee
which is derived in Appendix \ref{sec:appendix}.
In Ref.\cite{Pospelov_2008}, it was found that vector freeze-in through
kinetic mixing could account for the full DM abundance 
for $\epsilon \sim 10^{-11}-10^{-12}$ over the $\sim$ keV-MeV mass range. 
However, from \Eq{eq:kinmix}, it is clear that
the induced kinetic mixing can easily be subdominant
to dipole production through the operator in \Eq{eq:lint}; throughout this paper, 
we will only consider parameter space for which this requirement holds.


\subsection{Leptonic Couplings}
\label{sec:leptons}

In this section we consider the decay channels that 
arise from coupling $\Ap$ to charged leptons 
with $f = \ell$ in \Eq{eq:lint}, where $\ell = e,\mu,\tau$ is the flavor of the dipole interaction.
For $m_{\Ap} > 2m_\ell$, the dominant decay channel is $\Ap \to \ell \bar \ell$, which is generically too prompt for the dark photon
to serve as a viable dark matter candidate. However, for $m_{\Ap} < 2 m_\ell$, the $\Ap \to 3 \gamma$ channel shown at the top of Fig. \ref{fig:cartoon} can be cosmologically metastable due to phase-space suppression, so throughout this paper, we will only consider this mass ordering.
In the $m_{\Ap} \ll m_\ell$ limit, the $3\gamma$ decay width is 
\be
\label{eq:3gamma}
\Gamma_{\Ap \to 3\gamma} = \frac{\alpha^3 d_\ell^2 m_{\Ap}^9}{155520 \, \pi^4 m_\ell^6 } ~,
\ee
which corresponds to a vector lifetime of 
\be
\tau_{\Ap} \approx 10^{18} \, \text{Gyr}
\brac{10^{-10} \rm }{d_\ell \cdot \rm \, GeV}^2
\brac{10 \, \rm keV}{m_{\Ap}}^9 \brac{m_\ell}{m_e}^6  \!\!,~~~~~
\ee
so the $\Ap$ can easily be metastable on cosmological timescales if there are no faster decay channels. 

The $\Ap$ can also decay to neutrinos through
one-loop diagrams involving virtual $W$ exchange, as shown in the bottom of Fig. \ref{fig:cartoon}. The partial width for this
process is
\be
\label{eq:weakdecay}
\Gamma_{\Ap \to \bar \nu \nu} = \frac{d^2_\ell G_F^2 m_\ell^2 m_{\Ap}^5}{4 \pi^3} \log^2 \! \brac{m_W}{m_\ell}~,
\ee
and the ratio of partial widths satisfies
\be
\frac{\Gamma_{\Ap \to \bar \nu \nu}}{\Gamma_{\Ap \to 3\gamma}}
&\approx& 3 \times 10^{-3} \brac{m_\ell}{ m_e }^8 \brac{10 \, \rm keV}{m_{\Ap}}^4~,
\ee
where $G_F = 1.16 \times 10^{-2}$ GeV$^{-2}$ is the Fermi constant and we have set $m_\ell = m_e$ inside the log of \Eq{eq:weakdecay}. Since
 avoiding cosmologically prompt $\Ap \to \bar \ell \ell$ decays requires $m_{\Ap} < 2 m_\ell$, saturating this inequality maximizes the dominance of the photon 
 channel 
 \be
 \label{eq:comparison}
\frac{\Gamma_{\Ap \to \bar \nu \nu}}{\Gamma_{\Ap \to 3\gamma}} \approx
1.5 \times 10^{-2} \brac{m_\ell}{m_\mu}^4 \biggr |_{m_{\Ap} = 2m_\ell}  ~.
\ee
Thus, for $\ell = e$ or $\mu$ it is possible for the visible $3\gamma$ channel
to dominate over the $\Ap \to \bar \nu \nu$ channel while maintaining $m_{\Ap} < 2m_\ell$; for $\ell = \tau$ most $\Ap$ decays are invisible, but there can still be a subdominant photon signal from the $3\gamma$ decay.

\begin{figure}[t]
\center
\includegraphics[width=2.8in]{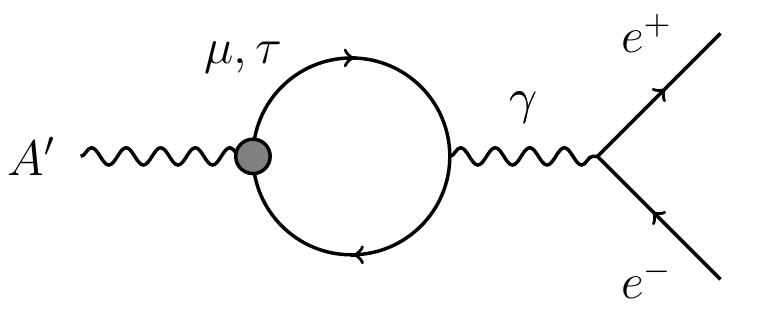}
\caption{Example Feynman diagram for an $\Ap\to e^+e^-$ decay through the induced kinetic
mixing in \Eq{eq:kinmix}. Although we only consider $m_{\Ap} < 2m_\ell$ so that
$\Ap \to \ell^+ \ell^-$ decays are forbidden, these loop-level two-body decays can dominate over the $3\gamma$ channel  
for $m_{\Ap} > 2m_e$. Similarly $\Ap \to \mu^+\mu^-$ decays can arise for $m_\Ap > 2m_\mu$ if $\Ap$ has a dipole
coupling to taus.}
\label{fig:FSR}
\end{figure}

Note that for $m_{\Ap} > 2m_{\ell^\prime}$, where $\ell^\prime$ is a lighter lepton flavor, there are also model dependent\footnote{Since kinetic mixing can receive ultraviolet contributions from heavy particle species beyond
the SM, the expression in \Eq{eq:kinmix} should be regarded as a representative lower limit.} $\Ap \to {\ell^\prime}^+ {\ell^\prime}^-$ decays induced by the kinetic mixing from \Eq{eq:kinmix}, as shown in Fig. \ref{fig:FSR}. The partial width for this process is 
\be
\Gamma_{\Ap \to {\ell^\prime}^+ {\ell^\prime}^-} = \frac{\epsilon^2 \alpha m_{\Ap}}{3} \left(1 + \frac{2m^2_{\ell^\prime}}{m^2_{\Ap}} \right)\sqrt{ 1 - \frac{4 m^2_{\ell^\prime}}{m^2_{\Ap}}}~.
\ee
 For $\ell = \mu$ and $\ell^\prime = e$, neglecting the phase space factors gives
 the ratio
\be
\frac{   
\Gamma_{\Ap \to {\ell^\prime}^+ {\ell^\prime}^-}
}{\Gamma_{\Ap \to 3\gamma}}  \approx 10^{10} \brac{m_\ell}{m_\mu}^8
\brac{50 \, \rm MeV}{m_{\Ap}}^8~,
\ee
so the kinetic mixing process is generically the dominant visible channel when kinetimatically accessible, and this can directly 
produce additional photons through final state radiation (FSR) via $\Ap \to {\ell^\prime}^+{\ell^\prime}^- \gamma$ decays. 

Although the total width properly includes the $3\gamma, \bar \nu\nu,$ and all available ${\ell^\prime}^+ {\ell^\prime}^-$ channels, in our analysis, we treat the kinetic mixing scenario separately as the magnitude of this mixing can vary considerably
depending on the full particle spectrum at high energies. In Figs. \ref{fig:mainfig} and \ref{fig:mainfigFSR}
we present our main results with and without kinetic mixing contributions, respectively (see Sec. \ref{sec:analysis} and \ref{sec:projections} for 
details).
 

\subsection{Hadronic Couplings}
\label{sec:quarks}
If the dipole in \Eq{eq:lint} involves SM quark fields and the vector mass is above the scale of QCD confinement, $m_{\Ap} \gg \Lambda_{\rm QCD} \approx 200$ MeV, many of the qualitative features from the leptonic scenario in Sec. \ref{sec:leptons} remain applicable. The key difference in the hadronic case is that the 2-body decay
through a $W$ loop yields two lighter quarks instead of neutrinos, so this 
channel may be observable even if the $3\gamma$ decay is subdominant. 

Since decay widths involving quark dipoles are parametrically similar to those involving leptons, the argument leading to \Eq{eq:comparison} remains valid for this scenario. Thus, for $m_{\Ap} < 2m_q$ and $m_{\Ap} > \Lambda_{\rm QCD}$, the loop level 2-body decay  will be dominant for all quarks with masses above the confinement scale ({\it i.e.} $q = c,t,b$). However,
for such heavier quark masses, it is generically difficult to ensure cosmological
metastability for dipole couplings that can accommodate the observed dark matter
abundance through freeze-in production (see Sec. \ref{sec:cosmo} below). 
For example,
using the loop induced 2-body width for $\Ap \to d \bar d$ through a charm quark 
dipole, we have
\be
\tau_{\Ap} \approx 3\times 10^{10} \, {\rm yr} \, 
\brac{10^{-14}}{ d_c \cdot \rm \, GeV}^2
\brac{m_{\Ap}}{500 \, \rm MeV}^5~,
\ee
where we have used the CKM matrix element $|V_{cd}| \approx 0.23$ \cite{ParticleDataGroup:2020ssz} for the 
$c\to d$ transitions in the electroweak loop. 
Thus, even for dipole couplings with GUT scale suppression, the $\Ap$ lifetime
is typically short for cosmological metastabiltiy.

In principle, it should be possible to evade this conclusion in
the $m_{\Ap} \ll \Lambda_{\rm QCD}$ regime where the smaller mass suppresses
the vector lifetime. However, in this scenario, the UV dipole coupling
to quarks must be matched onto the confined theory, which is beyond the scope of this work, but 
deserves further attention.
Thus, for the remainder of this paper, we will not consider quark dipole couplings 
in our numerical results.




\section{Cosmological Abundance }\label{sec:cosmo}

In this section we compute the $\Ap$ abundance through
freeze-in production. Since the dipole operator in \Eq{eq:lint} is not 
gauge invariant under $SU(2)_L\times U(1)_Y$, the dominant 
production processes will depend on whether the reheat temperature 
is above or below the scale of electroweak symmetry breaking, so we 
consider these cases separately.

Furthermore, since
the dipole interaction is a higher dimension operator, 
the $\Ap$ abundance is sensitive to the highest temperature
achieved in the early universe and the production is most efficient
at early times. Consequently, up to negligible corrections, 
the yields we calculate below are nearly identical for all lepton flavors and 
(up to color/charge factors) also apply to SM quarks if $T_{\rm RH} \gg \Lambda_{\rm QCD}$. 


\subsection{High Reheat Temperature $T_{\rm RH} > T_{\rm EW}$}

In the early universe, at temperatures above the scale of electroweak symmetry
breaking $T>T_{\rm EW} \approx 160$ GeV \cite{DOnofrio:2015gop}, the operator in \Eq{eq:lint} must be evaluated in its 
electroweak preserving form
\be
\label{eq:doublet}
{\cal L}_{\rm int} = \frac{ d_\ell }{ \sqrt{2} v} \, {\cal H} \, F^\prime_{\mu\nu} \bar L \sigma^{\mu\nu} \ell_R,
\ee
where $\cal H$ is the Higgs doublet, $L$ is a lepton doublet of any generation, 
$\ell_R$ is the corresponding right handed fermion, and 
$v = 246$ GeV is the Higgs vacuum expectation value. 
Assuming a negligible DM abundance
at reheating, the $\Ap$ population
arises predominantly from  pair annihilation 
$\ell^+\ell^- \to h \Ap$ 
and Compton-like production $\ell h \to \ell \Ap$
with corresponding cross sections
\be
\label{eq:cross-uv}
\sigma_{\ell^+\ell^- \to h \Ap} =  \frac{d_\ell^2 s}{48 \pi v^2}~~,~~
\sigma_{\ell h \to \ell \Ap} = \frac{d_\ell^2 s}{8 \pi v^2}~,~~~
\ee
where $s$ is the Mandelstam variable. Note that there are analogous processes 
involving the other doublet
components related by $SU(2)_L$ invariance
whose cross sections are equivalent to 
those in \Eq{eq:cross-uv}.

The thermally averaged cross section times velocity for these 
reactions can be written
\be
\langle \sigma v \rangle = \frac{1}{32T^5}
\int_0^\infty ds \, \sigma(s) s^{3/2} K_1\brac{\sqrt{s}}{T},
\ee
where $K_1$ is a modified Bessel function of the first kind
and we have taken the massless limit of the analogous
expression derived in Ref. \cite{Gondolo:1990dk}.

\begin{figure*}[t]
  \hspace{-0.8cm}
  \includegraphics[width=8.6cm]{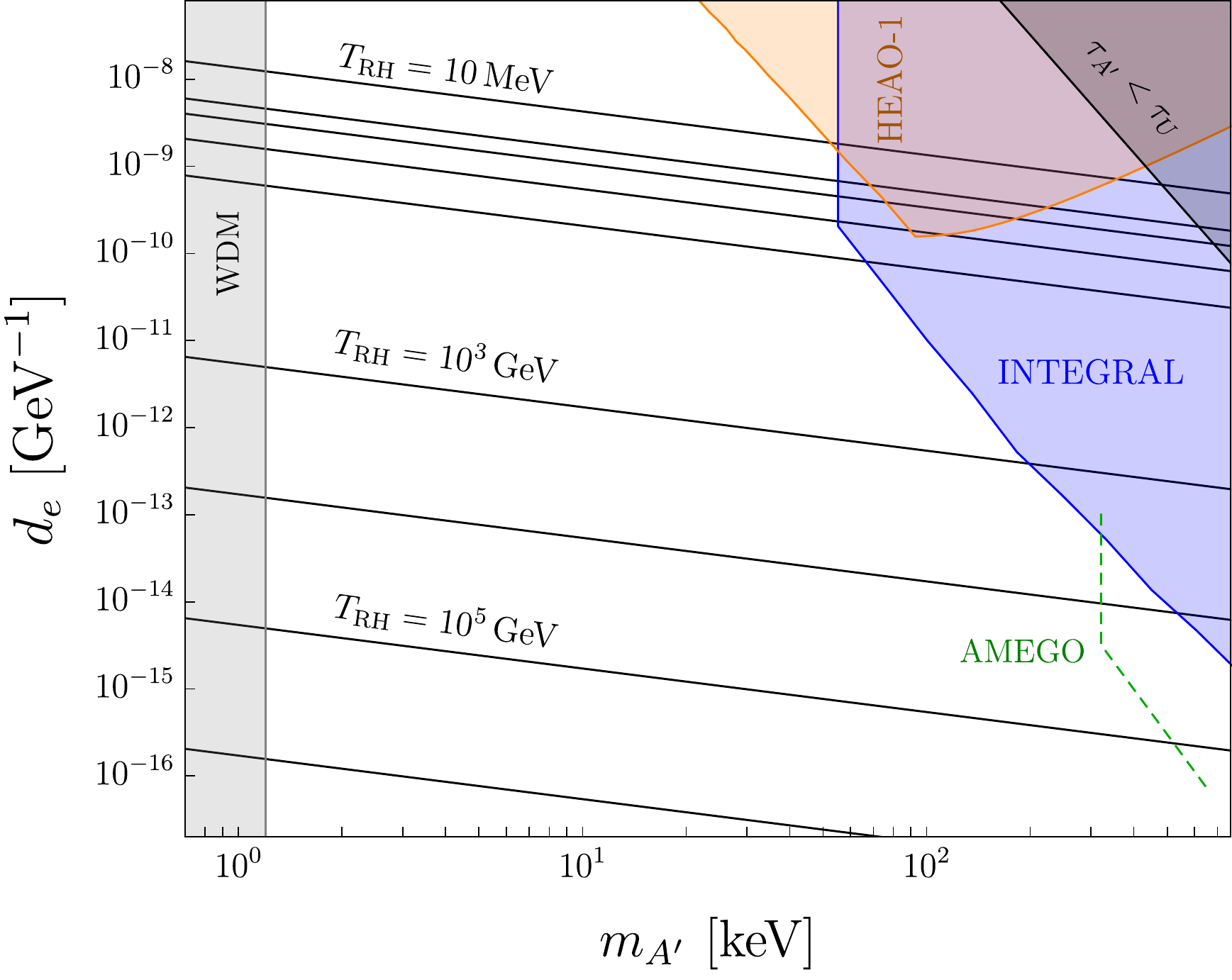}~~~~
  \includegraphics[width=8.6cm]{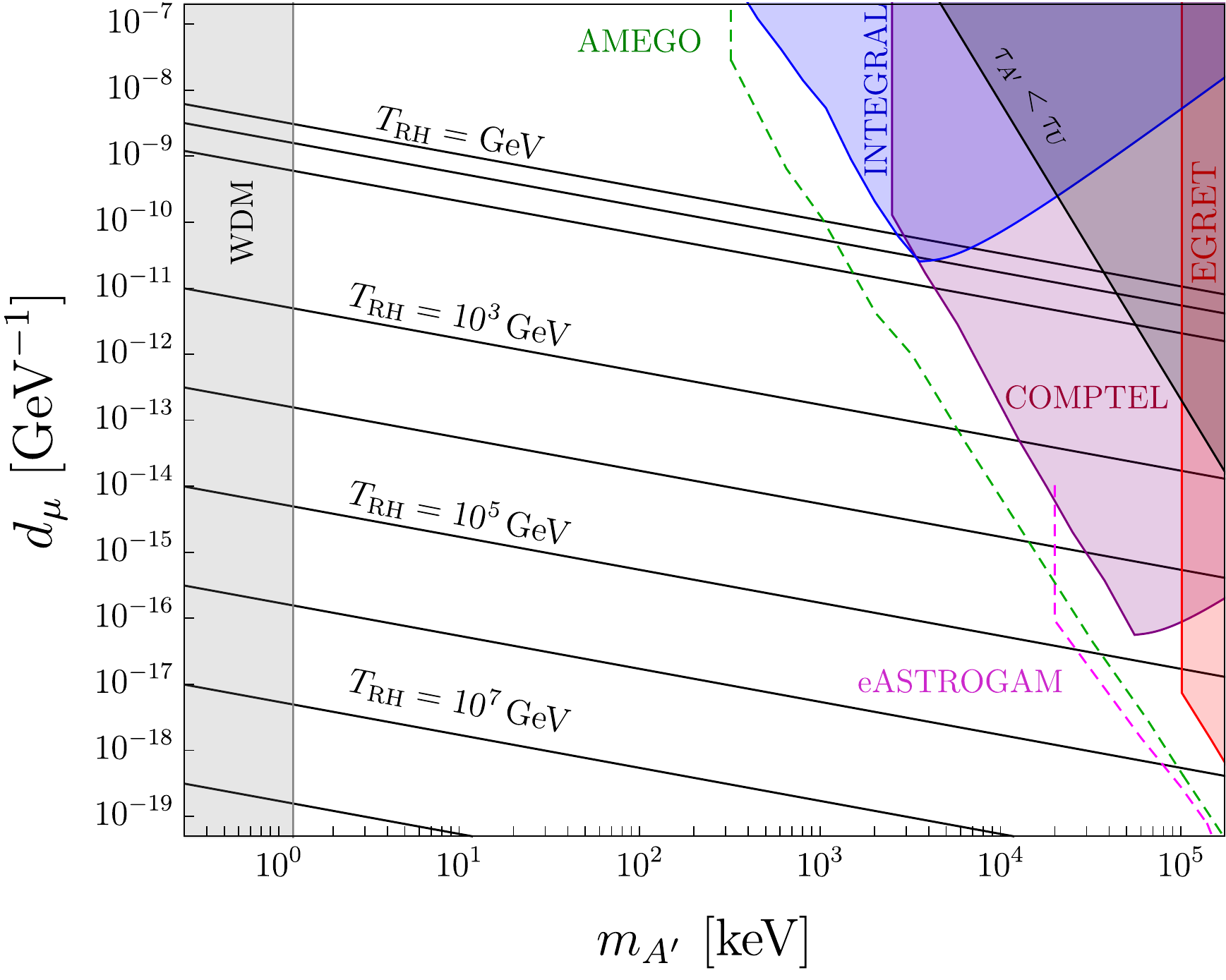} \\
  \hspace{-0.65cm}
    \includegraphics[width=8.6cm]{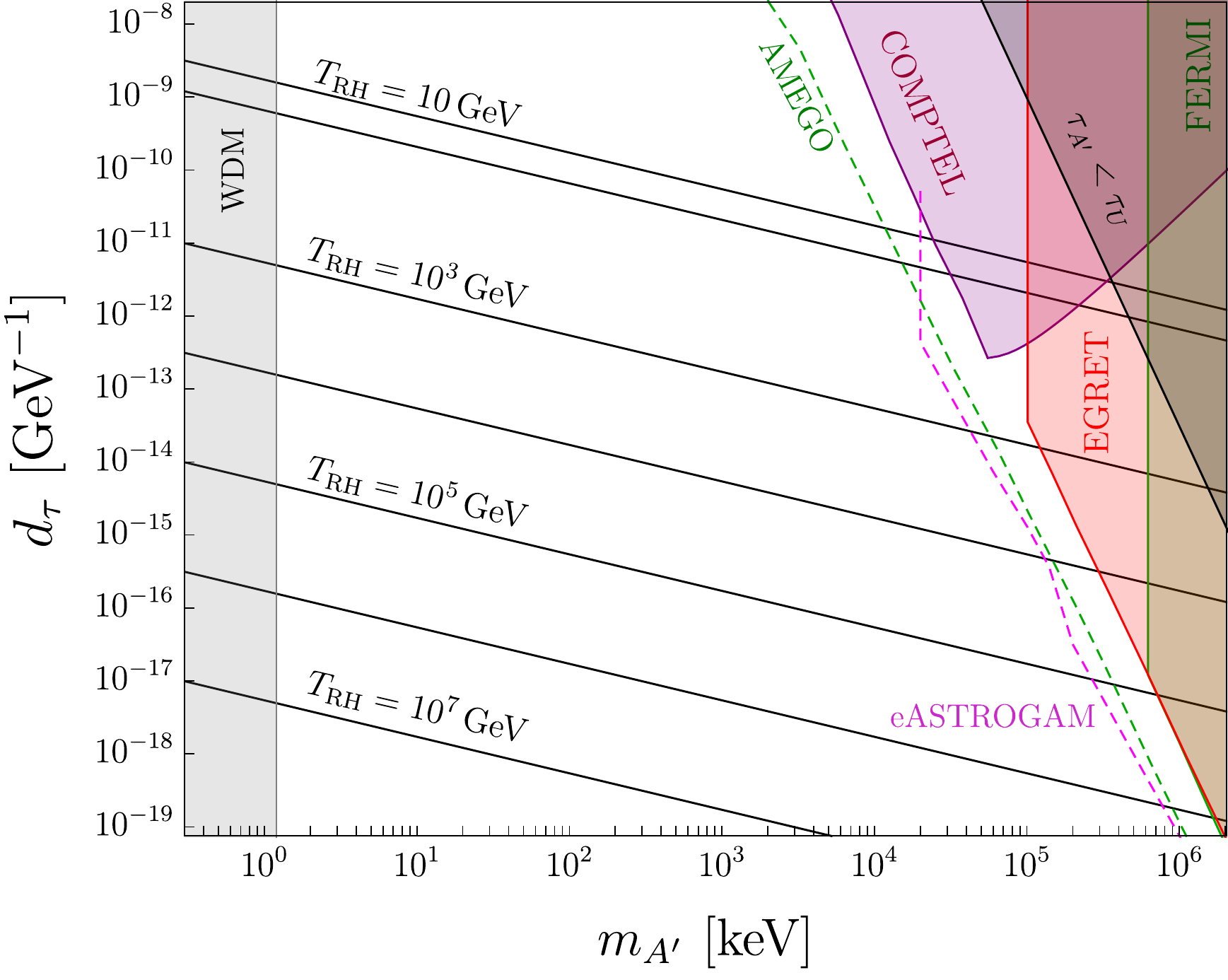} ~~~~~~
      \includegraphics[width=8.3cm]{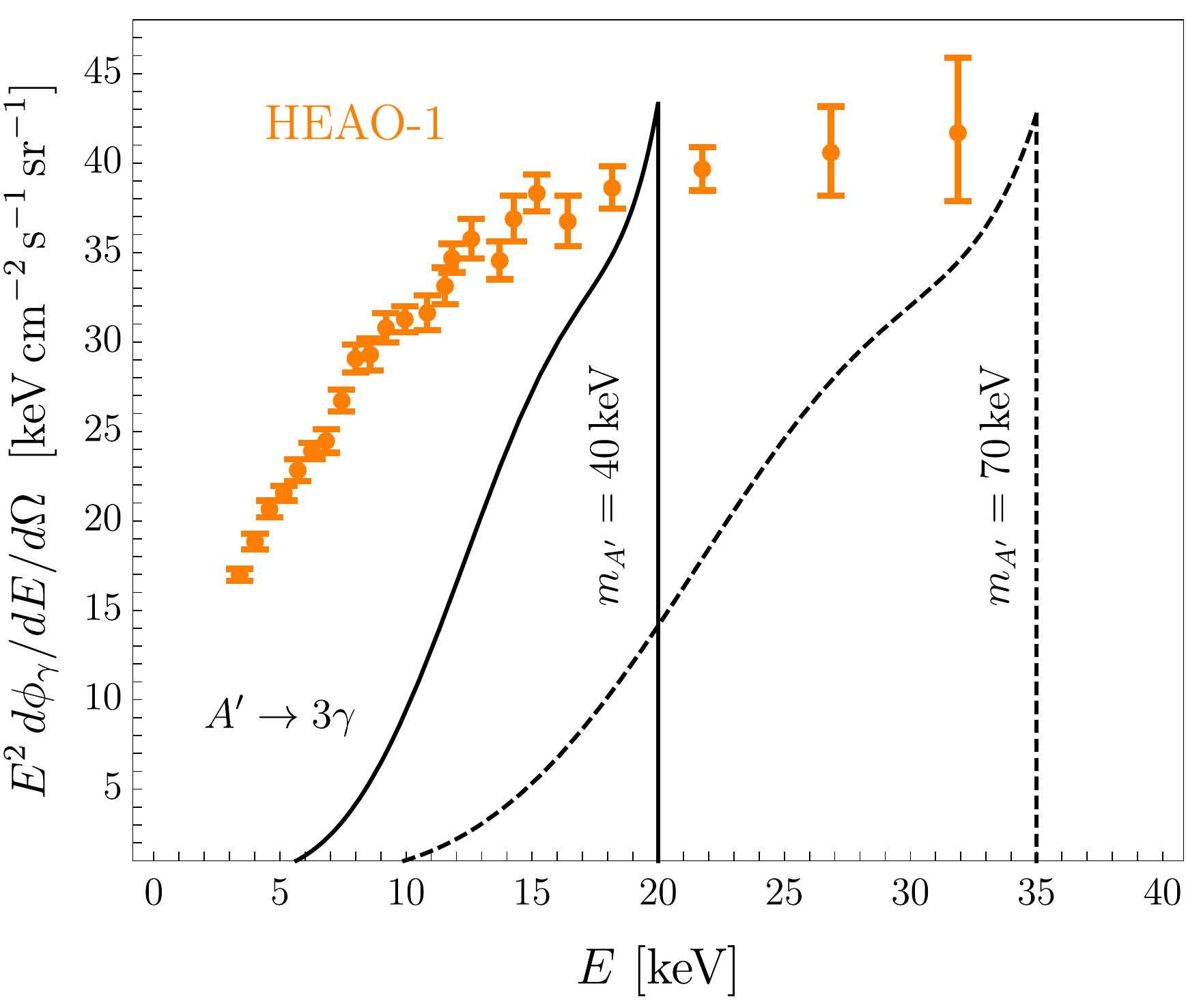}
  \caption{{\bf Top left:} parameter space for which $\Ap$ production via 
  dipole freeze-in achieves the observed DM abundance for various values of
  reheat temperature (black contours); note that few MeV is excluded by the successful predictions of standard BBN. 
  For higher values of reheat temperature, the abundance curves shift downwards by $\propto T^3_{\rm RH}$ from 
  \Eq{eq:abundance-uv}. Also shown are indirect
  detection limits on $\Ap \to 3 \gamma$
 from EGRET \cite{Strong:2004de,Strong:2003ey}, Fermi~\cite{Fermi-LAT:2012edv}, COMPTEL \cite{1994A&A...292...82S,Sreekumar:1997yg},
 INTEGRAL \cite{Bouchet:2008rp}, and HEAO-1 \cite{Gruber:1999yr}, and structure formation limits on warm dark matter (WDM) (see Sec. \ref{sec:structure}). We show future projections from e-ASTROGAM~\cite{e-ASTROGAM:2017pxr} and AMEGO~\cite{Kierans:2020otl}
 in dashed magenta and green curves, respectively. 
 The {\bf top right} and {\bf bottom left} panels show the same parameter space, but 
 for muon and tau couplings, respectively. Here we only include the $\Ap \to 3 \gamma$ channel and assume 
 no (model dependent) contributions from kinetic mixing which induces
 $\Ap \to e^+ e^-\gamma, \mu^+\mu^-\gamma$ decays that yield additional signal photons (see Fig. \ref{fig:mainfigFSR} for these additional contributions). Also note that the reheat temperature 
 is only evaluated above the SM lepton mass to ensure a relativstic population
 of such particles in the early universe.
 {\bf Bottom right:} flux comparison between the HEAO-1 X-ray (orange data points)
 \cite{Lumb:2002sw,DeLuca:2003eu,Nevalainen:2005iu,Hickox:2005dz,Carter:2007ng} and the predicted $\Ap \to 3 \gamma$ signal from \Eq{eq:flux}. We also show representative
  signals for $m_{A'}=40$~keV and $d_e = 5.7 \times 10^{-9} \, \rm GeV^{-1}$
 (solid black) 
  and $m_{\Ap} = 70$ keV with 
  $d_e = 4.6 \times 10^{-10} \, \rm GeV^{-1}$ (dashed black). }
  \label{fig:mainfig}
  \vspace{0cm}
\end{figure*}

In terms of the dimensionless yield $Y_{\Ap} = n_{\Ap}/{\cal S}$, where ${\cal S} = 2\pi^2 g_{\star,S} T^3/45$ is the 
entropy density and $g_{\star, S}$ is the number of entropic degrees of freedom, the Boltzmann equation
for $\Ap$ production can be written
\be
\label{eq:boltz}
\frac{dY_{\Ap}}{dT} = - \frac{4 n_\ell}{  H{\cal S}T}
\left[  
n_\ell \langle  \sigma v\rangle_{\ell\ell\to h \Ap} 
+ 2 n_h \langle \sigma v\rangle_{\ell h\to \ell\Ap}         \right],~~~
\ee
where $T$ is the photon temperature, $H = 1.66 \sqrt{g_{\star}} T^2/m_{\rm Pl}$ is the Hubble rate, $g_\star$ is the effective number of relativistic species,  $m_{\rm Pl}$ is the Planck mass, and $n_\ell  = 3 \zeta(3) T^3/(2\pi^2)$ and $n_h =  \zeta(3) T^3/\pi^2$ are the electron and Higgs number densities in
equilibrium;  we have neglected terms corresponding
to the reverse reactions ($\Ap \ell \to h \ell$ etc.) due
to the small relative $\Ap$ abundance in the early universe.
Note that the factor of 2 in the second term of \Eq{eq:boltz} 
accounts for Compton-like $\Ap$ production off both $\ell^\pm$
and the overall factor of 4 accounts for the multiplicity of states in the $\cal H$ doublet.

Since the dipole interaction is a higher dimension operator,  the $\Ap$ abundance is sensitive to the reheat
temperature of the universe, $T_{\rm RH}$. Assuming instantaneous
reheating and $g_\star = g_{\star, S} = $ constant throughout $\Ap$ production, \Eq{eq:boltz} can be integrated to 
obtain the asymptotic $\Ap$ yield at late times 
\be
Y^\infty_{\Ap} \approx 0.1 \, \frac{ d_\ell^2  \,T^3_{\rm RH} m_{\rm Pl} }{
g_{\star}^{3/2} v^2
}      ~~,
\ee
and the DM density fraction is 
 $\Omega_{\Ap} = m_{\Ap} s_0 Y_{\Ap}^\infty/\rho_c$, 
 where $s_0 = 2.1 \times 10^{-38}$ GeV$^{3}$ is the present day entropy density and 
$\rho_c = 4.1 \times 10^{-47}$ GeV$^4$ is the critical density.
Obtaining the observed DM abundance requires
an overall normalization 
\be
\label{eq:abundance-uv}
\Omega_{\Ap} 
&\approx& \Omega_{\rm DM} \brac{m_{\Ap}}{ 3 \, \rm MeV}
\brac{d_\ell \cdot \rm GeV}{10^{-13} }^2
\brac{T_{\rm RH}}{\rm TeV}^3
 \! ,~~~
\ee
which gives an adequate order of magnitude 
estimate. To obtain our final results, we numerically integrate 
\Eq{eq:boltz} to calculate $Y_{\Ap}^\infty$.
Note that our derivation is equally applicable to any lepton $\ell$ 
since the abundance is UV dominated and insensitive to the low-energy fermion mass
for all $T_{\rm RH} > T_{\rm EW}$.

Note that in the presence of nonzero kinetic mixing, there are
additional production channels through $f \bar f \to \Ap$ inverse decays, which 
can modify the cosmological $\Ap$ abundance. However,
we have verified that including this channel (and other production modes
that depend on the kinetic mixing), only contributes negligibly to the 
late time yield if the mixing arises from the dipole operator as 
in \Eq{eq:kinmix}.

\begin{figure*}[t]
  \hspace{-0.8cm}
\includegraphics[width=8.6cm]{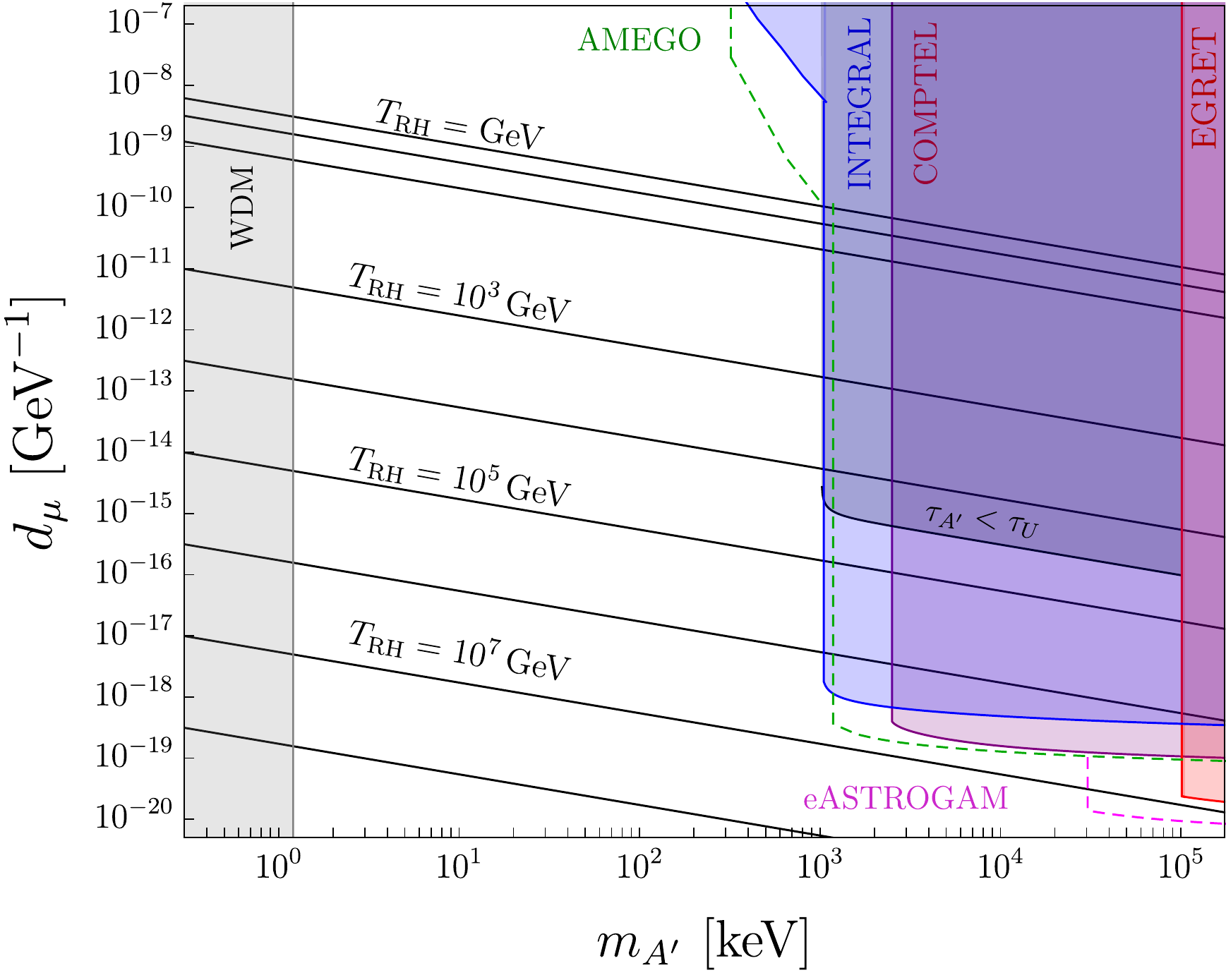}~~    
\includegraphics[width=8.6cm]{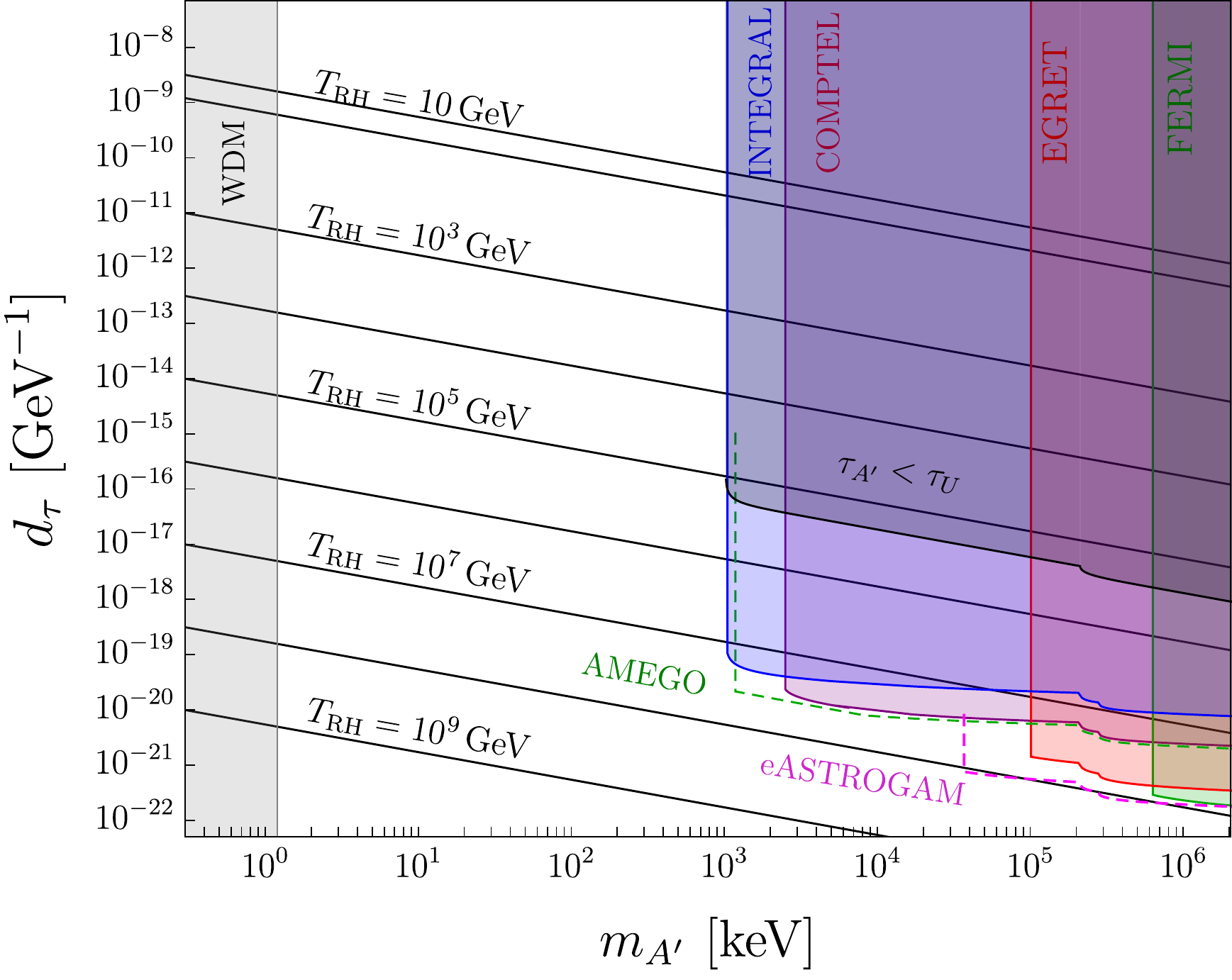} 
  \caption{ Same as the corresponding panels in Fig.~\ref{fig:mainfig},
  but including the effect of kinetic mixing for the muon and tau 
  flavor dipole couplings. Here the indirect detection signals
  are significantly affected by additional photons from
  $\Ap \to {\ell^\prime}^+{\ell^\prime}^- \gamma$ reactions where $\ell^\prime$,
  is a lighter lepton flavor than the tree-level dipole coupling to flavor 
  $\ell$
}
  \label{fig:mainfigFSR}
  \vspace{0cm}
\end{figure*}

\subsection{Low Reheat Temperature $T_{\rm RH} < T_{\rm EW}$}

If the reheat temperature is below the electroweak scale, 
the Higgs doublet is set to its vacuum expectation value 
$\langle H \rangle =  v/\sqrt{2}$ and \Eq{eq:doublet}
recovers \Eq{eq:lint}. In this regime, the
leading freeze-in reactions are $\ell^+\ell^-\to \gamma \Ap$
and $\ell^\pm \gamma \to \ell^\pm \Ap$ with respective cross sections
\be
\label{eq:sigmas-qed}
\sigma_{\ell\gamma  \to \ell \Ap} = \frac{\alpha d_\ell^2}{2} ~,~~~
\sigma_{\ell^+\ell^- \to \gamma \Ap} = \alpha d_\ell^2 ~~,~~
\ee
where $\alpha$ is the fine structure constant. 
The Boltzmann equation for $\Ap$ production now
becomes
\be
\label{eq:boltz2}
\frac{dY_{\Ap}}{dT} = - \frac{n_\ell}{ H {\cal S} T}
\left[  
n_\ell \langle  \sigma v\rangle_{\ell\ell\to \gamma \Ap} 
+ 2 n_\gamma \langle \sigma v\rangle_{\ell \gamma \to \ell\Ap}         \right],~~~
\ee
where $n_\gamma = 2 \zeta(2) T^3/\pi^2$ is the
photon number density in equilibrium and the thermal averages are trivial since
the cross sections in \Eq{eq:sigmas-qed} are constant
for $T_{\rm RH} \ll d_\ell^{-1}$, 
so $\langle \sigma v \rangle  \approx \sigma$ for these
processes. Integrating \Eq{eq:boltz2} from $T_{\rm RH} \to m_\ell$ and
approximating $g_\star = g_{\star, S} = $ constant, the
asymptotic yield is 
\be
Y_{\Ap}^\infty \approx 0.1 \, \alpha g_\star^{-3/2} d_\ell^2 T_{\rm RH} \, m_{\rm Pl}  ~~,
\ee
which corresponds to a present-day dark matter abundance of 
\be
\Omega_{\Ap} \approx \Omega_{\rm DM} \brac{d_\ell \cdot \rm GeV}{10^{-10}}^2 
\brac{m_{\Ap}}{\rm MeV} \brac{T_{\rm RH}}{\rm GeV},
\ee
where we have evaluated $g_\star$ at $T_{\rm RH}$. In our numerical 
results, we integrate the full expression in \Eq{eq:boltz2} to 
compute $\Ap$ the abundance.

\subsection{Inflationary Production}

In addition to the freeze-in abundance computed above, if the 
$\Ap$ has a nonzero mass during inflation, there is also an 
irreducible vector population produced through inflationary 
fluctuations \cite{Graham:2015rva}
\be
\Omega_{\Ap}^{\rm inf} \approx \Omega_{\rm DM} \sqrt{ \frac{m_{\Ap}}{6  \, \mu \rm eV} }
\brac{H_I}{10^{14} \, \rm GeV}^2~,
\ee
where $H_I$ is the Hubble rate during inflation. In our scenario, the {\it minimum}
Hubble rate during inflation satisfies $H_{I, \rm min} \sim T_{\rm RH}^2/m_{\rm Pl}$,
corresponding to an instantaneous transfer of energy from the inflaton to the 
SM radiation bath, so the minimum $\Ap$ abundance from 
inflationary production is 
\be
\Omega^{\rm min}_{\Ap} \approx 10^{-19} 
\sqrt{ \frac{m_{\Ap}}{ \rm MeV} }
\brac{T_{\rm RH}}{10^{10} \, \rm GeV}^4~,
\ee
which is negligible across our entire parameter space of interest.
Thus, assuming instantaneous reheating, if $T_{\rm RH}$ is sufficiently large for a non-trivial inflationary abundance, freeze-in production from \Eq{eq:abundance-uv} generically overcloses the universe.

If there is a large hierarchy between $H_I$ and $T_{\rm RH}$ (e.g. due 
an alternative cosmic expansion history \cite{Allahverdi:2020bys}), then freeze-in production
can be subdominant to inflationary production. However, independently of $T_{\rm RH}$, generating a non-trivial abundance generically requires a large value of $H_I \sim 10^{14}$ GeV, in some tension
with Planck limits on primordial tensor modes in CMB data \cite{Planck:2018vyg}. Thus, for 
the remainder of this work, we remain agnostic about the value of $H_I$ and neglect
any possible contribution from inflationary production, but it might be interesting 
to explore the full parameter space of such a hybrid scenario in future work.




\section{Structure Formation} \label{sec:structure}

In our scenario, the $\Ap$ population is mainly produced relativistically through 
freeze-in at high temperatures, near $T_{\text{RH}}$. For low values of 
$m_{\Ap} \sim$ keV, its phase space distribution can be warm at late times and 
erase small scale cosmological structure in conflict with various   
structure formation probes, including gravitational lensing, the
Ly-alpha forest, and the inventory of dwarf satellites in the Milky Way, among others. 

Constraints on warm dark matter (WDM) are typically calculated for thermal relics and assume that all of the
DM inherited a thermal velocity distribution at freeze out, in analogy with relic neutrinos.
Such constraints can also be applied to
feebly interacting DM particles that were never in equilibrium, but produced instead via freeze-in if
their velocity distribution has a nearly thermal profile. This is the case when the reaction rate is maximal near the freeze-in temperature
$T_{\text{FI}}$ at which most DM particles are produced. In our scenario, this production is dominated by reactions at $T_{\text{FI}} \sim T_{\text{RH}}$, as discussed in Sec. \ref{sec:cosmo}.

For WDM with a thermal spectrum, the constraint from structure formation can be calculated using the  free streaming length 
\be
    \ell_{\text{FS}} \equiv \int_{z_f}^{z_i} v(z) \frac{dz}{H(z)},\quad v(z) = \frac{p(z)}{\sqrt{m^2 + p^2(z)}},
\ee
where  $v(z)$ is the particle velocity, $p_i$ is its momentum at initial redshift $z_i$, and we have defined
\be
p(z) \equiv \brac{1+z}{1+z_i} p_i~.
\ee
Note that $\ell_{\rm FS}$ a monotonically growing function of  $p_{\text{com}}/m$, where 
\be
p_{\text{com}} = \frac{ p_i}{1+z_i}~,
\ee
is the comoving momentum of the particle, so the physical constraint on $\ell_{\text{FS}}$ can be translated into a constraint on the quantity 
\be
\frac{ p_{\text{com}}}{m} = \frac{p_i}{m (1+z_i)} \sim \frac{T_{\text{FI}}}{m (1+z_i)}~,
\ee
where $m$ is the mass of a thermal WDM candidate. 
In the absence of any entropy transfers into the primordial plasma, the quantity $T_{\text{FI}}/ (1+z_i)$ is constant and therefore, $\ell_{\rm FS}$ can be used to directly constrain the mass of the DM particle. However, since our scenario is sensitive to potentially high values of the reheat temperature, all entropy transfers at $T < T_{\rm RH}$ must be taken into account in order to translate $\ell_{\text{FS}}$ into a limit on the DM mass. 
Using entropy conservation, the $T/(1+z)$ ratio for our scenario relative to that of thermal WDM is given by 
\be  
\frac{T_{\Ap}}{T_\text{relic}} 
 \brac{1+z_{\text{relic}} }{1+z_{\Ap}} 
=
\left[ \frac{g_{\star}(T_\text{ relic})}{g_{\star}(T_{\text{RH}})}
    \right]^{1/3}~,
\ee
where $z_{\rm relic}$ is the redshift at which a thermal relic freezes out; this 
ratio be used to translate conventional WDM bounds into a lower limits on our dark photon mass.

In the literature (see e.g.\cite{Zelko:2022tgf} and refs therein), there are many different constraints on the mass of the thermal relic WDM particles extracted using different analysis methods. 
Such studies typically assume that DM freezes decouples from the SM at $T_\text{relic}^{\text{th}} \sim 2 \text{ MeV}$, so $g_{\star}(T_\text{relic}^{\text{th}}) = 10.75$. However, our dark photons are produced at $T_{\text{RH}}$ and if we take $g_{\star}(T_{\text{RH}}) = 106.75$ -- the total number of relativistic SM degrees of freedom at high temperature --  we conclude that
for the same $\ell_{\text{FS}}$, the analogous constraint on the dark photon mass is approximately $(106.75/10.75)^{1/3} \approx 2.15$ times weaker than traditionally reported limits on WDM thermal relics.

Although there are many such WDM limits in the literature (see Ref. \cite{Irsic:2017ixq} for a discussion),
we place conservative limits on our scenario using the weakest bounds from Ref. \cite{Garzilli:2018jqh}
which constrains $m_{\rm WDM} > 2.5$ keV by considering a  wider class of viable reionization models relative to other analyses.
Translating this limit into a bound on our scenario resuts in a constraint of $m_{A'} > 1.2$~keV. Note that this bound can be further relaxed if new particles with masses below $T_{\text{RH}}$ are thermalized in the early universe and provide additional entropy transfers into the SM radiation bath, resulting in  a larger value of $g_{\star}(T_{\text{RH}})$.


\section{Indirect Detection}
\label{sec:xray}


\subsection{General Formalism}
\label{sec:gen-form}

The differential photon flux from decaying DM in our our Galaxy is given by
\begin{equation}
\label{eq:flux}
    \frac{d\phi_{\gamma}}{dE d\Omega}
    =
    \frac{r_{\odot}}{4\pi}
    \frac{\rho_{\odot} \Gamma }{m_{A'}} 
    \frac{dN_{\gamma}}{dE} D,
\end{equation}
where $r_{\odot} =  8.5$~kpc is the solar distance from the Galactic center, $\rho_{\odot} =  0.3\text{ GeV} \text{cm}^{-3}$ is the local DM density, and $D$-factor is defined according to
\be
    D \equiv
\int_{l(\Omega)}
    \frac{dl}{r_{\odot}} 
     \frac{\rho(l,\Omega)}{\rho_{\odot}}~,
\ee
where the line integral is over the observed line of sight $l(\Omega)$ for a 
given solid angle $\Omega$. 
For the 3$\gamma$ decay channel, the inclusive single-photon spectrum is
\be
~~\frac{d\Gamma}{dE}
=
\dfrac{\alpha^3 d_\ell^2 m_{\Ap}^3 E^3}{9720\, \pi^4 m_\ell^6}  \left(35 m_{\Ap}^2  -130 E m_{\Ap} + 126 E^2 \right) \! ,~~~~~~~
\ee
where $E \le m_{\Ap}/2$ and the 
differential photon spectrum from three body decays is 
\be
    \frac{dN_{\gamma}}{dE}=\frac{3}{\Gamma } \frac{d\Gamma}{dE} ~,
\ee
where the factor of 3 accounts for the photon yield per decay event. Inserting this result into \Eq{eq:flux} alongside \Eq{eq:3gamma} yields our photon 
flux in terms of model parameters.

When the kinetic mixing in \Eq{eq:kinmix} is nonzero, for masses $m_{A'}> 2 m_{\ell^\prime}$, there are also $A' \to {\ell^\prime}^+ {\ell^\prime}^-$ decays, where ${\ell^\prime}$ is a fermionic species 
lighter than $\ell$, the original dipole flavor as depicted in Fig. \ref{fig:FSR}. 
These charged particles can yield potentially observable secondary photons via synchrotron radiation and inverse Compton scattering; such decays can also yield excesses in the cosmic positron spectrum. There are a number of works dedicated to constraining DM annihilation and decays into charged particles~\cite{Ibarra:2009dr,Boudaud:2016mos,Gaggero:2018zbd,Moskalenko:1999sx,Boudaud:2018oya}. However, in this work, we do not include this analysis. Instead, we present the conservative bounds from quantum corrections to the radiative tree-level process $A' \to {\ell^\prime}^+ {\ell^\prime}^- \gamma $ with an additional photon through FSR. The photon spectrum for this process can be written \cite{Siegert:2021upf}
\be
~~\frac{dN_\gamma}{dE}
\simeq 
\frac{\alpha[  m_{A'}^2 + (m_{A'}-2E)^2 ]}{2\pi \,m_{A'}^2 E} \log \left( \frac{m_{A'} (m_{A'}-2E)}{m_\ell^2} \right),~~~~~~
\ee
which arises by integrating the Altarelli-Parisi splitting function with a $\delta-$function. 


\subsection{Analysis Method}
\label{sec:analysis}

In this section we place limits on the signal from \Eq{eq:flux} using observations of the diffuse X-ray background from the HEAO-1~\cite{Gruber:1999yr}, INTEGRAL~\cite{Bouchet:2008rp}, COMPTEL~\cite{1994A&A...292...82S,Sreekumar:1997yg}, EGRET~\cite{Strong:2004de,Strong:2003ey} and Fermi~\cite{Fermi-LAT:2012edv} instruments. 
After removing known point sources from each data set, the resulting X-ray spectrum consists of three components: Galactic emission, instrumental backgrounds, and the diffuse X-ray background (XRB).
Since the Galactic and instrumental components contain a large number of spectral lines from atomic transitions, properly extracting the diffuse emission from the full spectrum requires a model of all relevant atomic lines and several additional power-law components. To properly constrain a  DM decay signal using this diffuse emission, such  modeling must be repeated in the presence of the additional DM induced spectral component, which is beyond the scope of this work. 

\begin{figure}[t]
  \hspace{0.cm}
  \includegraphics[width=8.8cm]{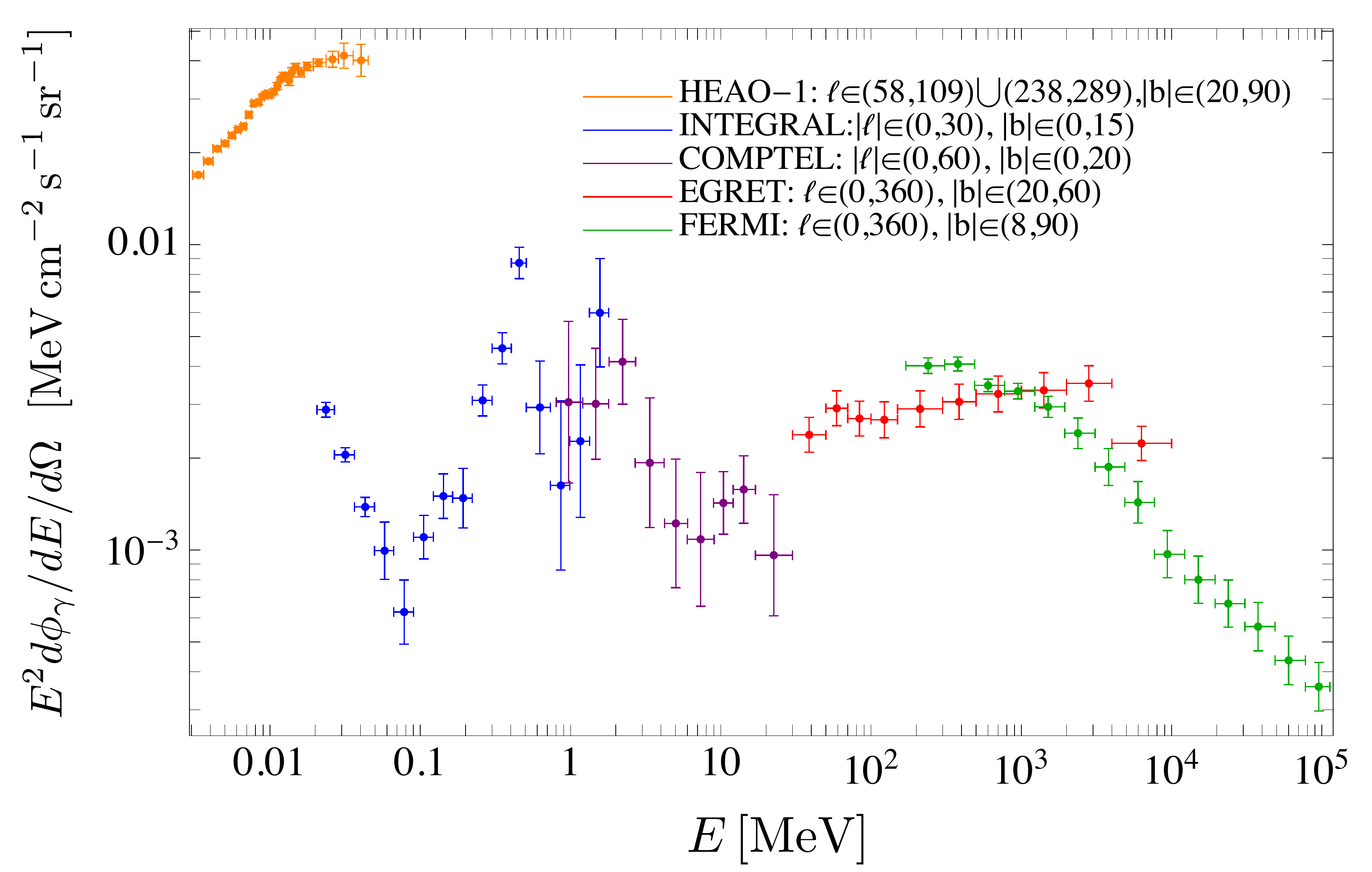}
  \caption{Observed X-ray flux data sets used to constrain the $\Ap \to 3\gamma$ signal in this scenario -- 
   figure adapted from Ref. \cite{Essig:2013goa}.
  Here the $\ell$ and $b$ values represent longitude and latitude coordinates within each instrument's field of
   view. Note that each instrument observes a different region of the sky, so the data sets presented here should
   not be compared against each other. }
  \label{fig:data}
  \vspace{0cm}
\end{figure}

 Instead, to extract a conservative, order of magnitude constraint from these instruments, we use the observational data shown in Fig.~\ref{fig:data} and, for each choice of DM mass, we demand that the number of signal photons in each bin does not exceed the number predicted by the XRB central value by more than two statistical standard deviations, except for the EGRET and Fermi, where the dominant systematic uncertainties are taken (see Ref. \cite{Essig:2013goa} for a discussion of this approach). For the data sets we use Galactic $D$-factors computed in Ref. \cite{Essig:2013goa}, which assumes a Navarro-Frenk-White DM profile \cite{NFW}, and our results for different $\Ap$ lepton dipole couplings are presented in Fig~\ref{fig:mainfig}.  In principle, the XMM-Newton telescope \cite{Carter:2007ng} can also be used to constrain this model, but 
we have verified that this limit corresponds to parameter space for which the freeze-in density can only be achieved for 
$T_{\rm RH} \ll $ MeV, which is not shown in Fig.~\ref{fig:data}.

\subsection{Future Projections}
\label{sec:projections}

In this section we compute projections for future missions with sensitivity to the $\Ap \to 3\gamma$ and  $\Ap \to {\ell^\prime}^+ {\ell^\prime}^- \gamma$ decay channels. We consider 
the next generation X-ray telescope Athena\footnote{http://www.the-athena-x-ray-observatory.eu} alongside MeV telescopes e-ASTROGAM~\cite{e-ASTROGAM:2017pxr} and AMEGO~\cite{Kierans:2020otl}, which can improve sensitivity to our decay signature.
Collectively, these future probes will have improved energy resolution, larger effective area, and wider fields of view, which  serve to
reduce astrophysical uncertainties in the background and improve signal reach. 

To model the sensitivity of these instruments, we bin our predicted signal in units of the reported energy resolution
for each telescope and demand that the visible DM decay signal not exceed the statistical uncertainty on
the photon background. Thus, for a minimal detectable flux in the $i^\text{th}$ energy bin, demanding 2$\sigma$ sensitivity,
$N_{\text{signal}} = 2 \sqrt{N_{\text{bg}}}$, yields 
\begin{equation}
    \int_{  E_{\min,i}}^{ E_{\max,i} } \frac{d\phi_{\text{signal}}}{dE} dE  = 2 \sqrt{\frac{d\phi_{\text{bg}}}{dE}\frac{\Delta E}{A_{\rm eff} \Delta t_{\rm obs}}}~~,
    \label{eq:Fmin}
\end{equation}
where $\Delta t_{\rm obs}$ is the observation time, 
$A_{\rm eff}(E)$ is the instrument's effective area, $\Delta E$ is the energy resolution, $d\phi_{\text{bg}}/dE$
is the background flux, and we integrate the signal over the energy range $(E_{\min,i}, E_{\max,i})$ spanned by the bin.

For our background flux estimates, we adopt XMM-Newton's models to compute Athena projections, COMPTEL and EGRET for e-ASTROGAM\footnote{We use the projected performance of e-ASTROGAM from Table 1.3.2 in paper~\cite{e-ASTROGAM:2017pxr}.}, and INTEGRAL, COMPTEL, EGRET and Fermi as proxies for our AMEGO projections. To calculate the relevant $D$-factor, we need to know the spatial orientations of these future telescopes, which are not yet finalized. Thus, for Athena, we use the average $D$-factor of the XMM-Newton blank sky background and re-scale for Athena's larger projected field of view. For e-ASTROGAM and AMEGO, we similarly re-scale the $D$-factors from INTEGRAL, COMPTEL, EGRET and Fermi correspondingly. 
In Fig.~\ref{fig:mainfig} we present our results including sensitivity projections for e-ASTROGAM and AMEGO;
 we find that Athena is only sensitive to dipole values which require $T_{\rm RH} \ll$ MeV to yield the observed freeze-in abundance, so its projections are not shown in Figs. \ref{fig:mainfig} and \ref{fig:mainfigFSR}. 
 

\section{Conclusion}
\label{sec:conclusion}

In this paper we introduced a simple model of vector DM with feeble magnetic dipole couplings to 
charged SM particles. In the early universe, the DM is produced non-thermally through freeze-in
and the present day abundance is sensitive to the reheat temperature, with different scaling 
before and after the electroweak phase transition.

If the vector mass is below the kinematic threshold for tree-level decays through 
the dipole coupling, loop level decays are generically suppressed either by phase space (for $\Ap \to 3 \gamma$) 
or by the $W$ mass for induced decays to lighter fermion species  (e.g. $\Ap \to \bar \nu \nu)$. Thus,  
for $m_{\Ap} < 2 m_f$ and dipole couplings that yield the observed DM abundance, the vector is generically metastable on cosmological timescales. If loop-induced kinetic mixing is also included, then for $m_{\Ap} > 2m_e$ there are additional
DM decay channels to pairs of charged particles throguh the electromagnetic current. 

For tree-level dipole couplings to leptons, the loop induced $\Ap \to 3\gamma$ decay and kinetic-mixing induced $\Ap \to e^+e^- \gamma, \mu^+ \mu^- \gamma$ decays predict visible photon signatures in the few-keV -- few-GeV energy range, where the lower limit is set by structure formation limits on WDM and the upper limit is set by the requirement that $m_{\Ap} < 2 m_\tau$ to avoid cosmologically prompt $\Ap \to \tau^+\tau^-$ decays if the $\Ap$ couples directly to taus. In this mass range, we have considered various observational constraints and computed projections for
future missions including the Athena, e-ASTROGAM, and AMEGO telescopes, which
will improve sensitivity to parameter space that yields the observed DM abundance
through freeze-in for various values of reheat temperature.

Although we have studied various indirect detection probes for our scenario, we 
note that there are several directions available for future work:

\begin{itemize}
    \item {\bf Charged Particle Decays:}
    In the presence of nonzero kinetic mixing, 
    the dominant decay channel for our DM candidate 
    is $\Ap  \to f^+f^-$, whenever this is kinematically available. 
    In our indirect detection analysis, we included signals from
    photons produced as FSR via $\Ap \to f^+f^- \gamma$, but neglected the possibility
    of secondary photons from the more common $\Ap \to f^+f^-$ process, which
    can yield additional detection handles from synchrotron radiation, inverse compton scattering, and antiparticle production. However, these channels require 
    dedicated modeling of astrophysical environments to extract signal predictions, which
    is beyond the scope of this paper.

    \item {\bf Quark Couplings:}
    If $\Ap$ couples
to quarks and $m_{\Ap} > \Lambda_{\rm QCD}$, its lifetime
is generically prompt for dipole couplings 
that can produce the observed DM abundance. For lighter $m_{\Ap} < \Lambda_{\rm QCD}$
it may be possible for a quark-coupled $\Ap$ to be a viable DM candidate, but 
investigating this mass range requires matching the $\Ap$-quark dipole operator onto
corresponding hadronic interactions below the QCD confinement scale, 
which we leave for future work.

\item {\bf Direct Detection: }
Finally, we note that this model may be testable at low mass direct detection experiments 
via $\Ap$ absorption onto detector targets, in analogy with searches for 
kinetically-mixed dark photon and axion-like dark matter candidates. Performing such a study
would require a reanalysis of existing bounds and future reach projections using a matrix
element for the dipole operator in \Eq{eq:lint}, which we also leave for future work.

\end{itemize}

\bigskip

\begin{acknowledgments}
We thank  Bogdan Dobrescu, Dan Hooper, Simon Knapen, Denys Malyshev, Elena Pinetti, Maxim Pospelov, and Tanner Trickle for helpful conversations. 
This work was performed in part at the Aspen Center for Physics, which is supported by National Science Foundation grant PHY-1607611. Fermilab is operated by Fermi Research Alliance, LLC, under Contract No. DE-AC02-07CH11359 with the US Department of Energy. This work has been supported by the Kavli Institute for Cosmological Physics at the University of Chicago through an endowment from the Kavli Foundation and its founder Fred Kavli.
\end{acknowledgments}

\begin{appendix}
\label{sec:appendix}
\section{Induced Kinetic Mixing}

The dipole coupling in \Eq{eq:lint} induces a kinetic mixing interaction between the dark and visible photon. At some high energy scale $\Lambda$ in the theory, the kinetic mixing amplitude is identically zero. This requires us to introduce the renormalization condition
\be
    \Pi^{\mu\nu} ( k^2 \to \Lambda^2) = 0~~,
\ee 
where $k$ is the momentum associated with this diagram. The leading order contribution to kinetic mixing can be written
\be
i\Pi^{\mu \nu} &=& i e d_f  k_\rho \intd{p} \frac{ \text{Tr} \Big[\sigma^{\mu \rho} (\slashed p - \slashed k + m_f) \gamma^{\nu} (\slashed p + m_f) \Big] }{[(p-k)^2 - m_f^2][p^2 - m_f^2]}, \nonumber
\ee
so using dimensional regularization and the modified minimal subtraction scheme, this integral becomes
\be
i \Pi^{\mu\nu} = \frac{i  e d_f m_f}{(4\pi)^2} \left( k^2 g^{\mu \nu} - k^\mu k^\nu \right) \int_0^1 dx \log \left[ \frac{m_f^2 -x(1-x) k^2 }{m_f^2 -x ( 1-x) \Lambda^2} \right]~. \nonumber
\ee 
In the limit $\Lambda^2 \gg m_f^2, k^2$, this integral takes the form
\begin{align}
    i \Pi^{\mu\nu} &= \frac{i  e d_f m_f}{8 \pi^2} \left( k^2 g^{\mu \nu} - k^\mu k^\nu \right)  \log\Big(\frac{m_f^2}{\Lambda^2}\Big)
\end{align}In order to express as an effective kinetic mixing coupling $\frac{ \epsilon}{2} F^{\mu \nu} F'_{\mu\nu}$, we simply remove the projector to obtain
,\begin{align}
    \epsilon = \frac{ e d_f m_f}{4 \pi^2}   \log\left(\frac{m_f^2}{\Lambda^2}\right)~,
\end{align}
which justifies the approximate form presented in \Eq{eq:kinmix}.

\end{appendix}

\bibliography{dipole}

\end{document}